\documentclass[12pt]{article}      
\usepackage[dvips]{graphicx,color} \usepackage{epsfig}
\DeclareGraphicsExtensions{.ep.gz,.eps,.ps,lps.gz}
\newcommand{\omm}{\Omega_{m0}}
\newcommand{\omlam}{\Omega_{\Lambda}}
\newcommand{\omq}{\Omega_{Q0}}

\newcommand{\bc}{\begin{center}}
\newcommand{\ec}{\end{center}}
\newcommand{\rat}{\mathcal R_0}
\begin{document}         
\title{\bf VACUUM ENERGY: ``IF NOT NOW, THEN WHEN?''}

\author{Sidney Bludman\\Deutsches Elektronen-Synchrotron DESY, Hamburg\\
  University of Pennsylvania, Philadelphia \thanks{Supported in
    part by U.S. Department of Energy grant DE-FGO2-95ER40893.}}

\maketitle

\abstract{For a flat universe presently dominated by static or dynamic
  vacuum energy, cosmological constant (LCDM) or quintessence (QCDM),
  we calculate the asymptotic collapsed mass fraction as function of
  the present ratio of smooth energy to matter energy $\rat=(1-\omm)/\omm$.
  Identifying these collapsed fractions as anthropic probabilities, we
  find the observed present ratio $\rat \sim 2$ to be likely in LCDM,
  but most likely in QCDM.}
\section {A Cosmological Constant or Quintessence?}

Absent a known symmetry principle protecting its value, no theoretical
reason for making the cosmological constant zero or small has been
found.  Inflation makes the universe flat, so that, at present, the
vacuum or smooth energy density $\Omega_{Q0}=1-\omm < 1$, is $10^{120}$
times smaller than would be expected on current particle theories.  To
explain this small but non-vanishing present value, a dynamic vacuum energy,
quintessence, has been invoked, which obeys the equation
of state $w_Q \equiv P/\rho <0$.  (The limiting case, $w_Q=-1$, a static
vacuum energy or Cosmological Constant, is homogeneous on all scales.)

The evidence for a flat low-density universe come from
\cite{WCOS,RHOR}: (1) The location of the first Doppler peak in the
CBR anisotroy at $l \sim 200$: $\Omega_m+\Omega_Q=1 \pm 0.2$; (2) The
slow evolution of rich clusters, the mass power spectrum, the CBR
anisotropy, the cosmic flow:$\omm=0.3 \pm 0.05$; (3) Curvature in the
SNIa Hubble diagram, dynamic age, height of first Doppler peak, cluster
evolution: $\omq=1-\omm \sim 2/3$.  Of these, the SNIa evidence is
most subject to systematic errors due to precursor intrinsic evolution
and the possibilty of grey dust extinction.  The combined data
nevertheless implies a flat, low-density universe with $\omm \sim 1/3$
and a smooth energy component with present energy density $\Omega_{Q0}
\sim 2/3$ and negative pressure $ -1 \leq w_Q \leq -1/2 $.

Accepting this small but non-vanishing value for static or dynamic
vacuum energy, a flat Friedmann cosmology (CDM) is characterized by
$\omm,~\Omega_{Q0}=1-\omm$ or the present ratio
 $$\mathcal R_0 \equiv u_0^3 \equiv \Omega_{Q0}/\omm=(1-\omm)/\omm~, $$
and by the equation of state for the smooth energy.
The {\em Cosmic Coincidence} problem now becomes
pressing: Why do we live when the clustered matter density
$\Omega(a)$, which is diluting as $a^{-3}$ with cosmic scale $a$, is
just now comparable to the static vacuum energy or present value of
the smooth energy i.e. when the ratio $\rat \sim 2$ ? 

In this paper, we distinguish the two limiting cases allowed
\cite{WCOS,RHOR} for the smooth energy component: LCDM: Cosmological
constant: $w_Q=-1$ and QCDM:\\ Quintessence: $ w_Q =-1/2$ .  In the
next section, we compare the expansion of these two limiting
low-density flat universes.  In Section 3, we extend to QCDM the
calculation of asymptotic mass fraction as function of a hypothetical
continuous variable $\omm$ presented by Martel {\it et
  al} \cite{MSW,MS} for QCDM.  Finally, we statistically infer that,
absent any prior information about $\omm$, the observed present ratio
$\rat$ is reasonable for a LCDM universe, and most likely for a QCDM
universe: ``If not now, then when?'' \cite{Hil}

\section {Expansion of a Low Density Flat Universe}

The Friedmann equation in a flat universe with clustered matter and
smooth energy density is
$$
H^2(x) \equiv (\dot{a}/a)^2=(8 \pi G/3)(\rho_m+\rho_Q), $$
or, in units of $\rho_{cr}(x)=3H^2(x)/8\pi G$,
$1=\Omega_m(x)+\Omega_Q(x),$ where the reciprocal scale factor $x \equiv
a_0/a \equiv 1+z \rightarrow \infty$ in the far past, $\rightarrow 0$
in the far future.

With the effective equation of state $w \equiv P/\rho=$ constant,
different kinds of energy density dilute at different rates $\rho \sim
a^{-n},~n \equiv 3(1+w)$, and contribute to
the deceleration at different rates $(1+3w)/2$ shown in the table:\\
   \begin{table}[h]
   \centering
   \begin{tabular*}{115mm}{@{\extracolsep{\fill}}l|ccc@{}}  \hline
   
   {\em substance}              &{\em w}       &{\em n}       &{\em (1+3w)/2}   \\  \hline
   radiation      &  1/3   &  4     &1    \\
   NR matter      &   0    &  3     &1/2      \\
   quintessence   & -1/2   &  3/2   &-1/4   \\
   cosmolconst    & -1     &  0     &-1     \\
   \hline  
   \end{tabular*}
   \caption{Energy Dilution for Various Equations of
       State}
   \end{table}\\
   The expansion rate in present Hubble units is
$$ H(x)/H_0=(\omm x^3+ (1-\omm) x^n_Q)^{1/2}. $$
The Friedmann equation has an unstable fixed point
in the far past and a stable attractor in the far future.  (Note the
tacit application of the anthropic principle: Why does our universe
expand, rather than contract?)
\begin{figure}[b]
\begin{center} 
\epsfig{file=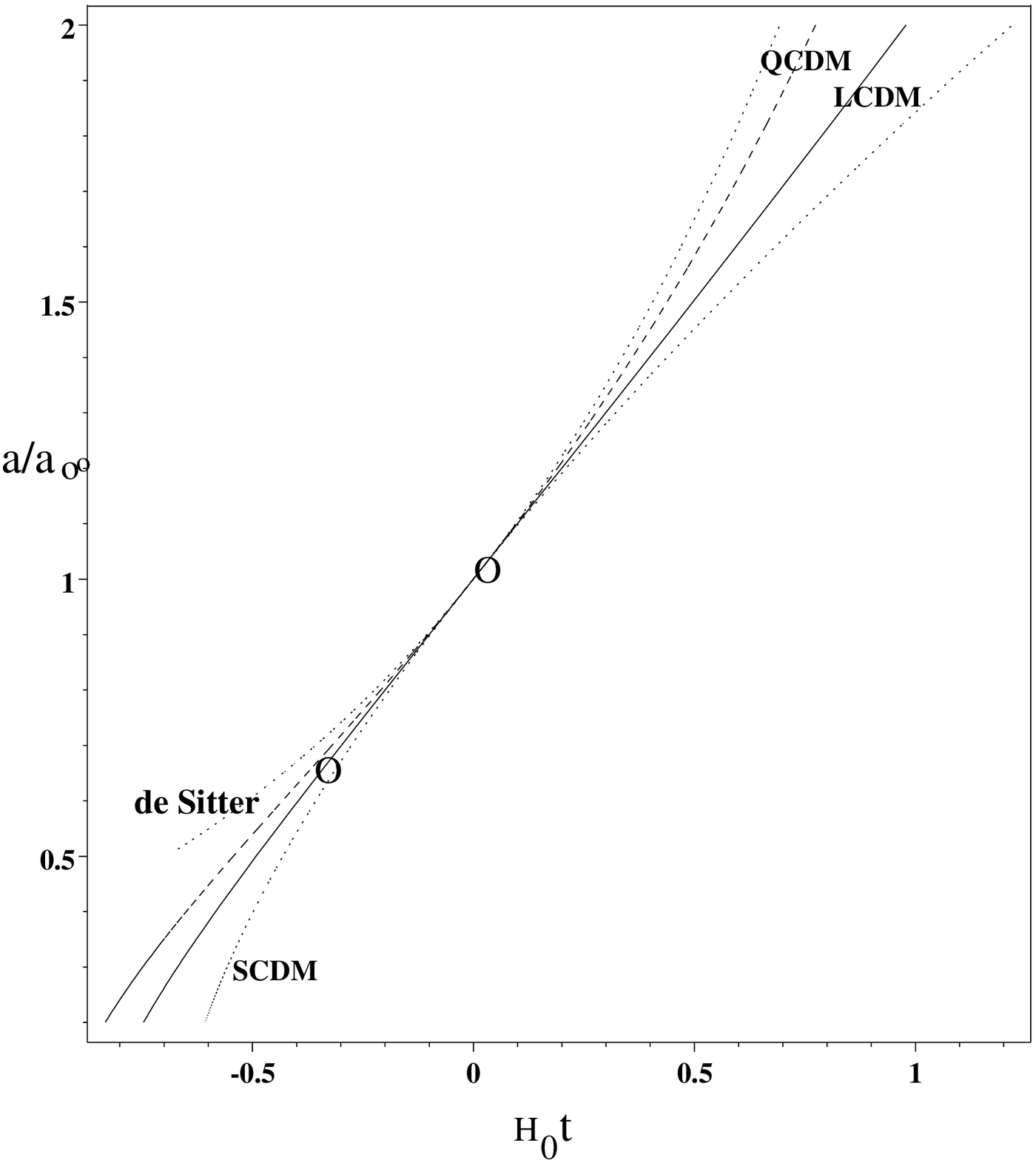,width=12cm,height=13cm} 
\caption{Scale evolution of LCDM and QCDM low-density flat universes
in the recent past and near future. The lower curve shows
the SCDM universe from which both LCDM and QCDM evolved in
the far past. The upper curve shows
the flat de Sitter universe towards which both LCDM and QCDM will evolve in
the far past. The inflection points marked (O) show where first LCDM
and later QCDM change over from decelerating to accelerating universes.}  
\end{center}      \end{figure}

The second Friedmann equation is $-\ddot{a} a/{\dot{a}^2}=(1+3w_Q
\Omega_Q)/2 $. 
The ratio of smooth energy to matter energy, $\mathcal R(a)=\mathcal
R_0 (a_0/a)^{3w_Q}$, 
increases as the cosmic expansion dilutes the matter density.
A flat universe, characterized by $R_0,~w_Q$, evolves out of an SCDM
universe in the remote past towards a flat de Sitter universe in the future.  
As shown by the inflection points (O) on the middle curves of Figure 1,
for fixed $\mathcal R_0$, QCDM  expands
faster than LCDM, but begins accelerating only
at the present epoch.  The top and bottom curves refer respectively to
a de Sitter universe ($\Omega_m=0$), which is always accelerating, and
an SCDM universe ($\Omega_m=1$), which is always decelerating.

The matter-smooth energy transition (``freeze-out'')
$\Omega_Q/\Omega_m=1$ took place only recently at $x*^{-w_Q}=\mathcal
R_0^{1/3} \equiv u_0$ or at $x*=1+z*=u_0^2=1.59$ for QCDM and, even
later, at $x*=1+z*=u_0=1.26$ for LCDM. Because, for the same value of
$u_0$, a matter-QCDM freeze-out would take place earlier and more
slowly than a matter-LCDM freeze-out, it imposes a stronger constraint
on structure evolution. 
As summarized in the table below, quintessence dominance begins 3.6
 Gyr earlier and more gradually than cosmological constant dominance.
 (In this table, the deceleration $q(x) \equiv -\ddot{a}/aH_0^2$ is
 measured in {\em present} Hubble units.)  The recent lookback time
$$H_0t_L(z)=z-(1+q_0)z^2+...,\quad z<1 ,$$
where $q_0=0$ for QCDM and $=- 1/2$ for LCDM.

   \begin{table}[h] 
   \centering
   \noindent
   \begin{tabular*}{125mm}{@{\extracolsep{\fill}}l|cc@{}}  \hline
   {\em event}                 &  {\em LCDM}      &  {\em QCDM} \\
   \hline \hline
   {\bf Onset of Vacuum Dominance}&               &                \\
   reciprocal scale x*=$a_0/a=1+z$                &$u_0$=1.260           &$u_0^2$=1.587     \\
   age   $t(x*)/H_0^{-1}$    &0.720             &0.478           \\
    in units $h_{65}^{-1}$Gyr      &10.8              &7.2            \\ \hline
   horizon size in units $cH_0^{-1}$   &2.39              &1.58           \\
    in units $h_{65}^{-1}$Gpc      &11.0              &7.24            \\ \hline
   deceleration q(x*) at freeze-out               &-0.333            &0.333
   \\  \hline \hline
   {\bf Present Epoch}         &                  &              \\
   age $t_0/H_0^{-1}$         &0.936             &0.845          \\
         $h_{65}^{-1}$Gyr      &14.0              &12.7           \\ \hline
   horizon in units $cH_0^{-1}$                      &3.26              &2.96          \\
    in units $h_{65}^{-1}$Gpc &15.0              &13.6            \\ \hline
   present deceleration $q_0$  &-0.500            &0              \\ \hline

   \end{tabular*}
   \caption{ Comparative Evolution of LCDM and QCDM Universes}
   \end{table}

\section{Evolution of Large Scale Structure}

In this section, we extend to QCDM earlier LCDM calculations
\cite{MSW,GLV,MS} of the asymptotic mass fraction $f_{c,\infty}$ that
ultimately collapses into evolved galaxies.  This is presumably a
measure of the number density of galaxies like our own, that are
potentially habitable by intelligent life. 
We then compare the QCDM and LCDM asymptotic
mass fraction distribution functions, as function of an assumed
$\omm$.

The background density for large-scale structure formation
is overwhelmingly Cold Dark Matter (CDM), consisting of clustered
matter $\Omega_m$ and smooth energy or quintessence $\Omega_Q$.
Baryons, contributing only a fraction to $\Omega_m$, collapse after
the CDM and, particularly in small systems, produce the large
overdensities that we see.
  
Structure formation begins and ends with matter dominance, and is
characterized by two scales: The horizon scale at the first
cross-over, from radiation to matter dominance, determines the power
spectrum $P(k,a)$, which is presently characterized by a shape factor
$\Gamma_0=\omm h =0.25 \pm 0.05$.  The horizon scale at the second
cross-over, from matter to smooth energy, determines a second scale
factor, which for quintessence, is at $\sim 130$ Mpc, the scale of
voids and superclusters.  A cosmological constant is smooth at all
scales.
  
Quasars formed as far back as $z \sim 5$, galaxies at $z \geq 6.7$,
ionizing sources at $z=(10-30)$.  The formation of {\em any} such
structures, already sets an upper bound $x*<30$ or $(\omlam /
\omm)<1000, \omq<30$, for {\em any} structure to have formed.  A much
stronger upper bound, $u_0<5$, is set by when {\em typical} galaxies
form i.e. by estimating the {\em probability} of our observing
$\rat=2$ at the present epoch.

\subsection{Asymptotic Collapsed Mass Parameter $\beta$}  
Martel {\it et al} \cite{MSW} and Garriga {\it et al} \cite{MS} have
already calculated the asymptotic mass fraction from the
Press-Schechter formalism
$$f_{c,\infty} =\mbox{erfc}( \sqrt\beta
)=(2/\sqrt\pi)\int^{\infty}_{\sqrt\beta}\exp(-t^2)\,dt , $$
depending only on
$$\beta \equiv \delta_{i,c}^2/(2\sigma_i^2) , $$
where $\sigma_i^2$ is
the variance of the denssity field, smoothed on some scale $R_G$, and
$\delta_{i,c}$ is the minimum density contrast at recombination which
will ultimately make a bound structure.  This minimum density contrast
grows with scale factor $a$, and is, except for a numerical factor of
order unity \cite{MS}, $\delta_{i,c} \sim x*/(1+z_i)$.  Both numerator
and denominator in $\beta$ refer to the epoch of recombination, but
this factor $(1+z_i)$ cancels out in the quotient.  (MSW and MS have
improved on the Press-Schechter formalism by assuming spherical
collapse of Gaussian fluctuations or linear fluctuations that are
surrounded by equal volumes of compensating underdensity.  Except in
the limit $\beta \rightarrow 0$, the PS formula overestimates the
collapsed mass by factor $\approx (1.70)*\beta^{0.085}$, or about 40\%
near $\omm=1/3$.  For simplicitly, this paper adheres to the PS formula with
$R_G=1$ Mpc.  In a forthcoming paper, we will use the
improved MSW formula for both $R_G=1,~2$ Mpc.)

The variance of the mass power spectrum depends on the cosmological
model ($\omm$) and on the relevant co-moving galactic size scale
$R_G$, but is insensitive to $w_Q$, for $w_Q<-1/3$ \cite{WS}.  For the
QCDM model we consider, $\sigma_i^2(\omm,R_G)$ is therefore the same
as that already calculated \cite{MSW,MS} for LCDM, for a
scale-invariant mass spectrum smoothed with a top-hat window function.
For the observed ratio $\rat=2, \omm=1/3$, at recombination $1000
\sigma=3.5, 2.4$, for comoving galactic size scale $R_G=1,~2$ Mpc.

The numerical factor in $\delta_{i,c}$ is $9/5(4)^{1/3}=1.1339$ for
both $w_Q=-1$ and $w_Q=-1/2$, so that $\delta_{i,c}=1.1339
x*/(1+z_i)$.  Thus, the collapsed mass parameter
$\sqrt\beta=0.80x*/\sigma_i(R_G, u_0)$, depends explicitly on $u_0$
through $x*=u_0, u_0^2$ for LCDM, QCDM respectively. It also depends
implicitly on $u_0$ through $\sigma_i$.  Nevertheless, in going from LCDM
the argument of $f_{c,\infty}$ scales simply as
$\sqrt\beta_{QCDM}=\sqrt\beta_{LCDM}\cdot u_0$.

Both asymptotic mass fractions are practically unity for large $\omm$,
but fall off with increasing ratio $\mathcal R_0 >1$.    For any
$\rat>1$, QCDM always leads to a smaller asymptotic mass fraction than
LCDM. For ratio $\rat < 1$, $f_{c,\infty}$ changes slowly and
the differences between QCDM and LCDM are not large. At the observed
ratio $R_0=2$, the Press-Schechter asymptotic mass fractions are
$0.696,~0.623$ for LCDM, QCDM respectively.

\subsection{Asymptotic Collapsed Mass Fraction Distribution Function}
As function of the ratio $\omm$, the asymptotic mass fraction defines
a distribution function
$$f_{c,\infty}=d\mathcal P/d \rat .$$
In Figure 2, instead
of $f_{c,\infty}$ we plot the logarithmic distribution function in the
ratio $R_0$
$$F(\omm)=\rat \cdot f_{c,\infty}=d \mathcal P/d \log \rat , $$
for
LCDM and for QCDM and galactic size scale 1 Mpc. (Even for LCDM, this
differs by a factor $\sigma_i ^3(\omm)$ from the logarithmic
distribution in $\beta$, $d\mathcal P/d \log (\beta^{3/2})$ that is
plotted by MSW and GLV.) $F(\omm)$ may be thought of as the ratio
$R_0$ weighted by the number density of galaxies $f_{c,\infty}$.

The figure shows broad peaks in the logarithmic distributions in
$\omm$ at ($(\omm,F)=(0.23,1.27)$ for LCDM and at $(0.32,1.78)$ for
QCDM.  At the observed $\omm=1/3$, shown by circles (O), the
asymptotic mass fraction logarithmic distribution in $R_0$ falls at
97\% of the QCDM peak and at 78\% of the LCDM peak.
\begin{figure}[b]
\begin{center} 
\epsfig{file=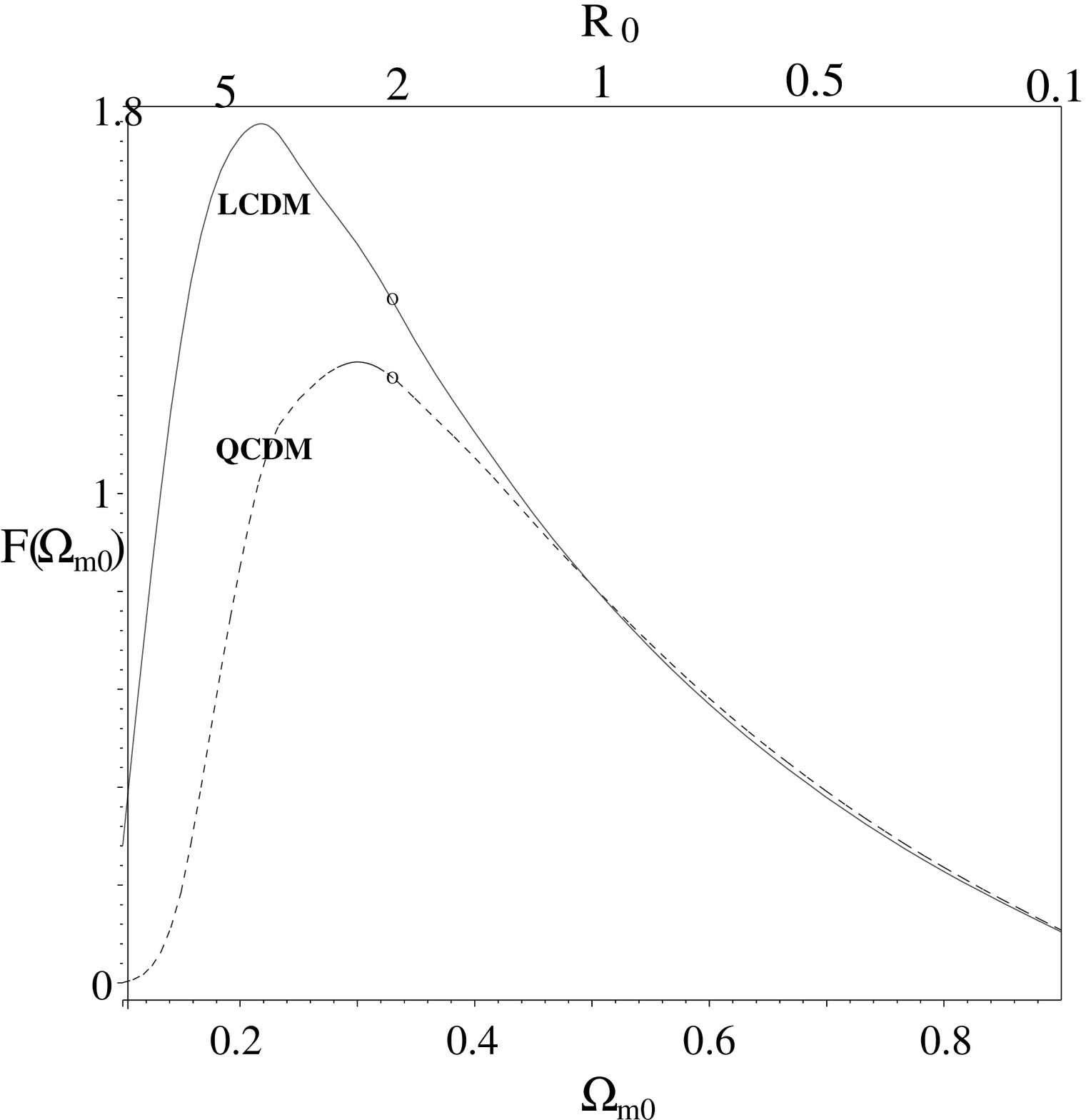,width=14cm,height=12cm}
\caption{Logarithmic distribution function for the Press-Schecter
  asymptotic collapsed mass fraction as function of hypothetical
  present matter density $\omm$ (bottom scale) or smooth energy/matter
  ratio $R_0$ (top scale). Our observed universe (O) with $\omm \sim
  1/3,~ R_0 \sim 2$ falls within the broad peak of the LCDM
  distribution and remarkably close to the peak of the QCDM
  distribution.}
\end{center}      \end{figure} 

\section{$\Omega_Q \sim \Omega_m$ is 
  Quite Likely for Our Universe}

It is not surprising that our universe, containing at least one
habitable galaxy, has $\rat=\mathcal O(1)$. What is impressive is that
our observed low-density universe, is almost exactly that which will
maximize the number of habitable galaxies.  Our existence does not
explain $\omm$, but the observed value makes our existence (and that
of other evolved galaxies) most likely.  

What epistemological
inference should we draw from this remarkable coincidence between our
observed universe and the possible asymptotic mass fractions in either
LCDM or QCDM universes?  What should we infer statistically about any
fundamental theory determining the parameters of our universe?
 
An anthropic interpretation has already been given \cite{E,V,MSW,GLV}
to ``explain'' a non-vanishing cosmological constant, in an assumed
universe of subuniverses with all possible values for the vacuum
energy $\omlam=1-\omm$.  In each of these subuniverses, the probability for
habitable galaxies to have emerged before the present epoch, is a
function of $\omlam$:
$$\mathcal Prob(\omm) \propto (\mbox{prior distribution in} ~\omm)
\times F(\omm). $$
As always, the overall probability depends on the assumed prior.
MSW, assuming nothing about initial conditions, take a prior flat in
$\omm$.  GLV argue that the prior should be determined by a theory
of initial conditions and is {\em not} flat for most theories.

We prefer not to assume a distribution of real subuniverses, but to
inversely apply Bayes' Theorem to our own universe.  In the absence of
any {\em physical} explanation of the smooth energy, or until one is
found, the partial information that intelligent astronomers exist
tells us the observed smooth energy is just about what would be
expected from equal {\it a priori} probabilities for $\omm$ in the
very early universe.  That our universe is realized at or near the
maximum in the asymptotic mass distribution function confirms that the
prior is flat or peaked at $\omm \sim 1/3$. Any phenomenologically
viable fundamental theory must ultimately produce this value
or be indifferent to the cosmological parameters.

This research benefited from useful discussions with H. Martel and
with M. Roos.


\begin{thebibliography}{99}

\bibitem{WCOS} L. Wang, R.R. Caldwell, J.P. Ostriker and
  P.J. Steinhardt, astro-ph/9901388.

\bibitem{RHOR} M. Roos and Haron-or-Rashid, astro-ph/003040.

\bibitem{Hil} Rabbi Hillel, Mishnah, PirqeiAvot 1.14 (c. 20 A.D.).

\bibitem{E} G. Efstathiou, M.N.R.A.S. {\bf 274}, L73 (1995).

\bibitem{V} A. Vilenkin, Phys. Rev. Lett. {\bf 74}, 846 (1995).

\bibitem{MSW} H. Martel, P.R. Shapiro and S. Weinberg, Ap. J. {\bf
    492},29 (1998) (MSW).

\bibitem{GLV} J. Garriga, M. Livio and A. Vilenkin, astro-ph/9906210 (GLV).

\bibitem{WS} L. Wang and P.J. Steinhardt, Ap. J {\bf 508}, 483 (1998).

\bibitem{MS} H. Martel and P.R. Shapiro, astro-ph/9903425 (MS).
\end{thebibliography}
\end{document}